**Precise determination of the saturation coverage of polygons *in silico* using exclusion assisted packing technique**

Aref Abbasi Moud[1]*

[1] Department of Chemical and Petroleum Engineering, University of Calgary, 2500 University Dr NW, 5 Calgary, AB, T2N 1N4, Canada

*Author to whom correspondence should be addressed; electronic mail: aabbasim@ucalgary.ca

**Abstract:** To fill the allotted area without overlapping, a regular or asymmetric filling of an empty space must be created using the time-dependent packing technique known as random sequential adsorption (RSA). Reaching saturation is a difficult undertaking since the density of coverage tends to hit a limit in the infinite-time limit. In this study, we try to precisely determine the saturated packing of oriented 2-d polygons, such as squares (4-sided), regular convex polygons such as pentagons (5-sided), hexagons (6-sided), heptagons (7-sided), octagons (8-sided), nonagons (9-sided), decagons (10-sided), dodecagons (12-sided), and polygon with 15 (Pentadecagon) and 20 sides (Icosagon) using a newly developed algorithm named here exclusion assisted packing. While using a different technique, our findings are in line with earlier extrapolation-based research as well as more recent attempts to establish strictly saturated packaging. We applied the "separating axis theorem" to determine if freshly inserted and previously placed polygons overlap. In these circumstances, we show that saturation as a lower limit is achieved when the RSA addition slows down too much. The next step was using a newly developed technique to saturate the available space. The results of polygons' calculated saturation packing were found to be comparable to results obtained from other methods and techniques presented in the references (Cieśla et al., 2021; Zhang, 2018; Zhang & Torquato, 2013).

## 1. Introduction

There are several uses for random packings, a popular model for random media, in theoretical and practical situations(Lee & Subbiah, 1991; Torquato & Jiao, 2009; Truskett et al., 1998). Random close packing (RCP), a sort of random packing, is one where the packing is created continuously until the density stops rising and the particles are in touch with one another. Although RCP is frequently used to simulate granular materials(Jaeger & Nagel, 1992; Liu et al., 2018; Meng et al., 2016), its definition raises questions since it is challenging to maximize both packing density and disorder at the same time(Torquato et al., 2000). Additionally, the precise measurement techniques utilised, whether numerical or experimental, might have an impact on the density of the packing(Asencio et al., 2017).

RSA is another type of packing generation technique that has a well-defined mean packing density(Evans, 1993). A virtual particle is chosen at random for location and orientation inside the packing, and if it does not intersect with any other particles, it is added to the packing with its position and orientation fixed. This process is used to generate the packing. The virtual particle is eliminated if it does connect with other particles. Up to the point where there is no more room big enough to add another particle, this procedure is repeated. RSA was first developed early in the 20th century(Flory, 1953), but it gained widespread attention due to Feder's observation that RSA packings resemble monolayers created during irreversible adsorption processes(Feder, 1980).

The creation of experimental adsorption layers on surfaces or interfaces has been modeled in a variety of industrial contexts using RSA. As in the case of Pickering emulsions in the oil industry(Moud, 2022), these layers are frequently produced by molecules or particles adhering to a surface and forming one. Modeling these layers and examining the interactions between adsorbed particles and the surface, as well as the impact of these interactions on molecule orientation inside the layer, may be done using RSA algorithms, which generate two-dimensional loose random packings. Researchers can assess how well RSA predicts adsorption kinetics and the shape of the adsorption layer by comparing experimentally determined features with those derived from RSA modeling (Hemmersam et al., 2008; Kosior et al., 2020; Manzi et al., 2019; Min et al., 2017).

In addition to its practical applications, RSA also serves as a simple yet effective protocol for forming disordered packings that consider excluded volume effects(Baule, 2019; Baule, 2017; Onsager & Runnels, 1963). This makes it a useful tool for studying the properties of a wide range of different shapes in colloidal systems. However, it should be noted that RSA is not suitable for studying other phenomena induced by excluded volume, such as phase transitions(Frenkel, 1987; Onsager, 1949), as the system it produces is essentially non-equilibrium(Torquato, 2018).

Numerous investigations on RSA have included packings made up of mono-disperse items of different forms. Classical geometries like discs (Zhang & Torquato, 2013), squares(Brosilow et al., 1991), polygons (Zhang, 2018), and ellipses(Haiduk et al., 2018) have been used in these forms as well as more unusual ones such smoothed dimers(Cieśla et al., 2016), disco rectangles(Lebovka et al., 2011), star polygons(Cieśla & Barbasz, 2014), and polygons with rounded corners(Cieśla et al., 2021). In the one-dimensional case, the saturation packing fraction can be obtained analytically as 0.747597920 (Rényi, 1958) however for two-dimensional and 3-D the saturation packing fraction for discs and spheres has been estimated only through numerical simulations; the most precise ones are 0.5470735 ± 0.0000028 for two dimensional and 0.3841307 ± 0.000002 for 3-D cases(Zhang & Torquato, 2013). Other figures reported for two dimensional simulations of discs are 0.5470735(Zhang & Torquato, 2013), 0.547067(Cieśla & Ziff, 2018), 0.547070(Cieśla & Nowak, 2016), 0.5470690(Wang, 1994), 0.54700(Torquato et al., 2006), 0.54711(Chen & Holmes-Cerfon, 2017), 0.5472(Hinrichsen et al., 1990), 0.547(Feder & Giaever, 1980), 0.5479(Wang, 2000). The most accurate report for discs is 0.547068869 ± 0.000000057 and can be found in ref (Kubala et al., 2022). Similarly for 2-D aligned squares saturation coverage reported in the literature is 0.562009(Brosilow et al., 1991), 0.5623(Wang, 2000), 0.562(Feder & Giaever, 1980), 0.5565(Blaisdell & Solomon, 1970), 0.5625(Dickman et al., 1991), 0.5444(Tory et al., 1983), 0.5629(Akeda & Hori, 1976), 0.562(Jodrey & Tory, 1980).

One of the main challenges with RSA modeling is how the algorithm performs when it approaches saturation. In adsorption experiments, saturated monolayers may frequently be created in a couple of minutes; however, the conventional RSA strategy becomes unsuccessful when there is little chance of finding a large enough region for a second particle. This results in a large increase in the expected number of RSA iterations needed to find such a space, which lengthens simulation times. Additionally, because the algorithm is unable to determine when the packing is saturated, it may continue to try to add additional particles even after saturation is reached, even when doing so is no longer feasible.

Methods have been developed to address the problems with the efficiency of the RSA algorithm near saturation, but these methods have only been successful and reported only for specific shapes such as discs(Ebeida et al., 2012; Wang, 1994; Zhang & Torquato, 2013), ellipses, sphero-cylinders, and polygons(Kasperek et al., 2018) (Cieśla et al., 2019; Zhang, 2018).

This paper presents a method for generating RSA packings using polygons with fixed orientation and tests its effectiveness. The study begins with squares and ends with polygons with side counts of 12,15 and 20 and employs the RSA approach to calculate the saturation packing limit for convex polygons with up to 20 sides. The results showed that the samples gradually produced structures with increasing packing as the number of sides in the polygons increased, and values close to those observed for discs were regenerated. The "separating axis theorem" technique was used to determine if there was a collision between two polygons in this study, and the details of the algorithm are described in the following sections.

In the study of adsorption layers, basic geometric forms like discs, two-dimensional spherocylinders, or ellipses(Adamczyk, 2017; Matijevic & Borkovec, 2004) are frequently used to mimic the cross-section of complicated molecules. Numerous molecules, even ones with more complicated structures that may be modeled using numerous discs(Adamczyk et al., 2010; Cieśla et al., 2013), can be represented using these models. Nevertheless, the models made using this approach can have rough, concave cross-sections and unwanted steric effects. To analyze polygon adsorption in greater depth, it may thus be essential to assess more complicated cross sections, as will be presented here. Due to its simplicity and lower computing cost, the polygon in this instance has been chosen with a fixed orientation. In further experiments, we will use more stringent techniques, such as adding arbitraty orientation to polygons that may be able to more clearly capture molecular adoptions.

This work presents a novel approach to generate RSA packings using arbitrary polygons with a fixed orientation, allowing the construction of models that more accurately simulate the adsorption of practically any two-dimensional shape and more accurately represent the convex cross-sections of many molecules. Researchers may more precisely analyze the adsorption processes by employing these polygons, especially when it comes to the anchoring of orientated particles that may be caused by flow-induced orientation. This study also aims to investigate how the form of an adsorbed particle affects the primary characteristics of their random packing as measured by mean saturation packing.

## 2. Modeling and simulation techniques

Colloidal particles often diffuse near the surface during the adsorption process. This procedure may result in a film made of molecules that have been randomly adsorbed due to adhesion. Here, our attention is on irreversible adsorption that results in adsorbate monolayers. Molecular dynamics (MD) is a simple method for numerically simulating these events. The benefits of MD include precise prediction and control over most environmental variables, including temperature and the diffusion constant(Binder, 1995). The main issue is the performance fault, and the reference (Berendsen, 1999) addresses the limitations of MD techniques in adsorption applications. Due to this, we chose to employ a different technique called continuum RSA, which has been used to explore colloidal systems with success(Adamczyk et al., 1997).

To perform simulation:

- A virtual particle was produced (polygons with a varied number of sides), and based on a uniform probability distribution, its position in an area was picked at random.
- An overlap test (subject of next section) with a virtual particle's adsorbed previous closest neighbors was conducted. This test determines whether a particle's surface-to-surface distance is higher than zero.
- The virtual particle was adsorbed and added to an already-existing covering layer if there was no overlap.
- The virtual particle was eliminated and abandoned if there was an overlap.

## 2.1 Proposed algorithm: Exclusion assisted packing

The existence of an intersection between two polygons may be checked using a variety of procedures. The "separating axis theorem" is one such approach that uses mathematical equations to determine if two polygons are overlapping(Gottschalk et al., 1996). According to the separating axis theorem (SAT), two convex polygons do not cross if a line separates them. The normal to one of the edges of either polygon can be referred to as the separation axis, which is a line.

The following actions can be conducted to determine whether two polygons (here referred to as Polygon A and Polygon B) cross using the separating axis theorem:

- Calculate the edge normal for each edge in polygon A, then project both polygons onto the edge normal.
- Determine each polygon's minimum and maximum projection onto the normal.
- The polygons do not overlap if the maximum projection of polygon A is smaller than the minimum projection of polygon B, or vice versa.

- Repeat the process for each edge in polygon B if the projections overlap.
- The polygons connect if their projections onto all separating axes do not overlap.

While the SAT can be used to detect if two convex polygons overlap, it is possible that non-convex polygons or polygons with holes will not be covered by this theorem. Other methods could be required in certain circumstances to check for the intersection.

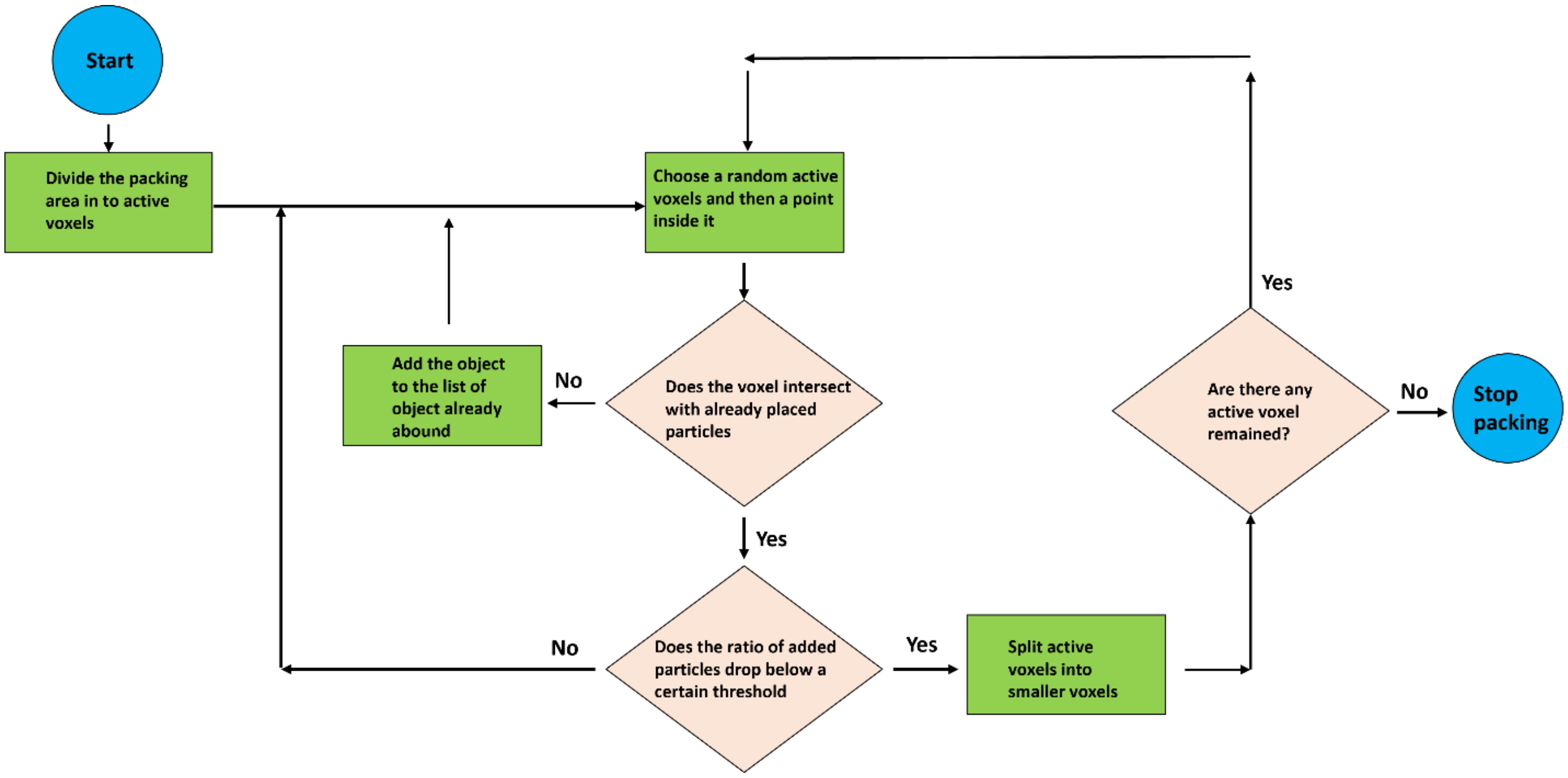

**Figure 1**. Block diagram of the algorithm that generates strictly saturated RSA packing.

For this study, in addition to the classical RSA model which we showcase qualitatively here, we pursued the application of another method that generates strictly saturated RSA packings; algorithms with similar foundation with their details have been explained comprehensively in two recent publications(Zhang, 2018; Zhang & Torquato, 2013). We name this method “exclusion-assisted packing”. To explain our algorithm, the initial active voxels are all the equally sized meshes that make up the collector area before the algorithm is launched. The next step is to randomly select a location inside the voxel and determine if placing a polygon there would cause it to collide with polygons already in place. If the response is "yes," the voxel is classified as inactive; if "no," a new particle is added to the pool of adsorbed particles. The details of the process are shown in **Figure 1**.

The initial mesh size chosen for the collector was 50 by 50 in a range of 1 million. This was eventually increased to a final mesh size of 0.015 units, which resulted in approximately 12 million voxels. The following conditions were used for the simulation:

- Boundary condition: Fixed
- For each polygon, the simulation was repeated 100 times and the average as mean saturation packing was reported.
- Both extrapolation and voxel elimination methods were used, and the results were compared.
- A total of 10 objects (squares, pentagons, hexagons, heptagons, octagons, decagons, dodecagons, 15-gons, and 20-gons) were assessed.

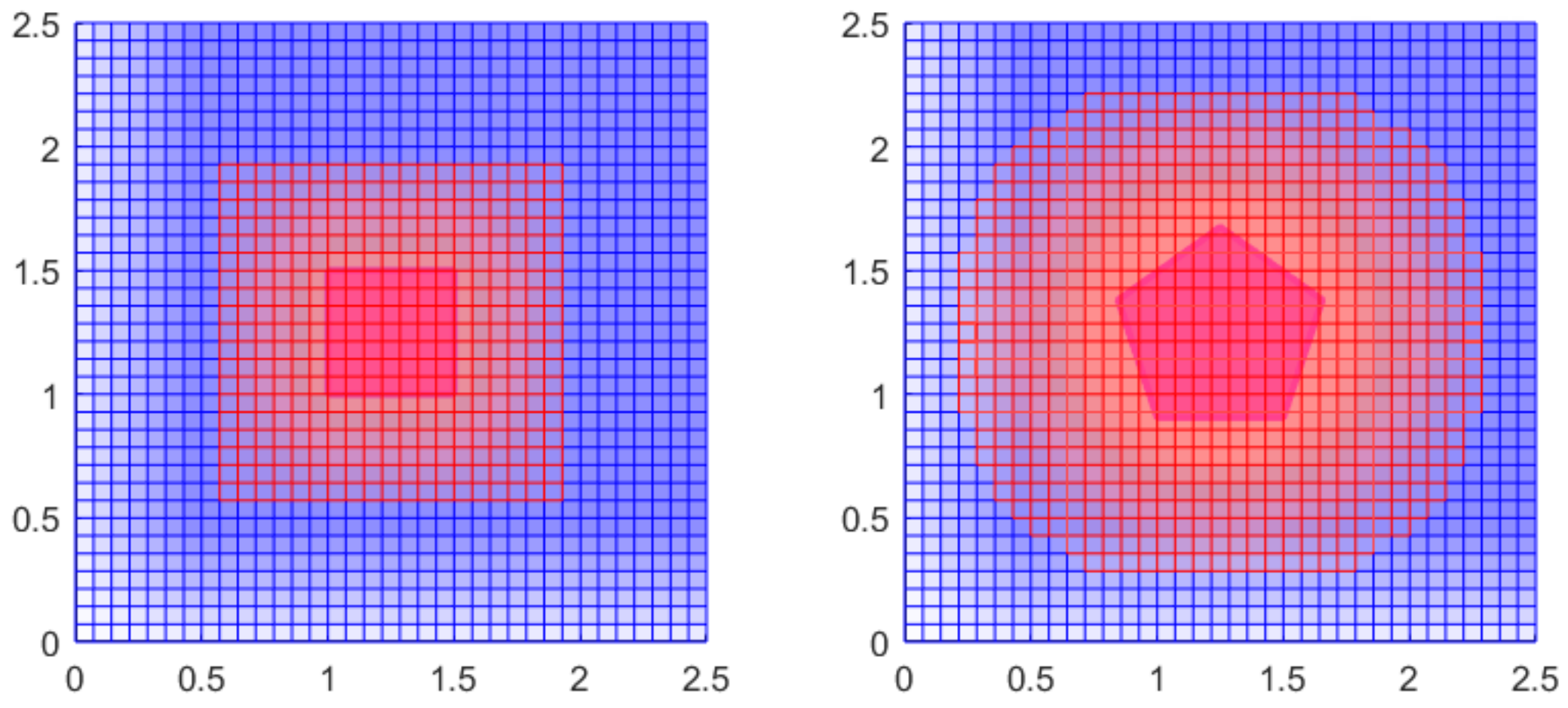

**Figure 2**. Exclusion zones in the algebraic approach **(a)** symbolic model square **(b)** symbolic model pentagon. The black frame is the particle. The red and blue areas correspond to exclusion and available zones for possible neighboring particles, respectively. The side length for both the square and pentagon is 0.5.

The eliminated voxels shown in the previous section reside primarily in the exclusion zone as shown in **Figure 2. Figure 2** has been used as an example to illustrate how the simulation process works; it shows how square and pentagonal shapes have had their locations segregated by square voxels. Due to the region's inaccessibility to other particles for the blue voxels in the block diagram, voxels with a red border are ineligible and deleted.

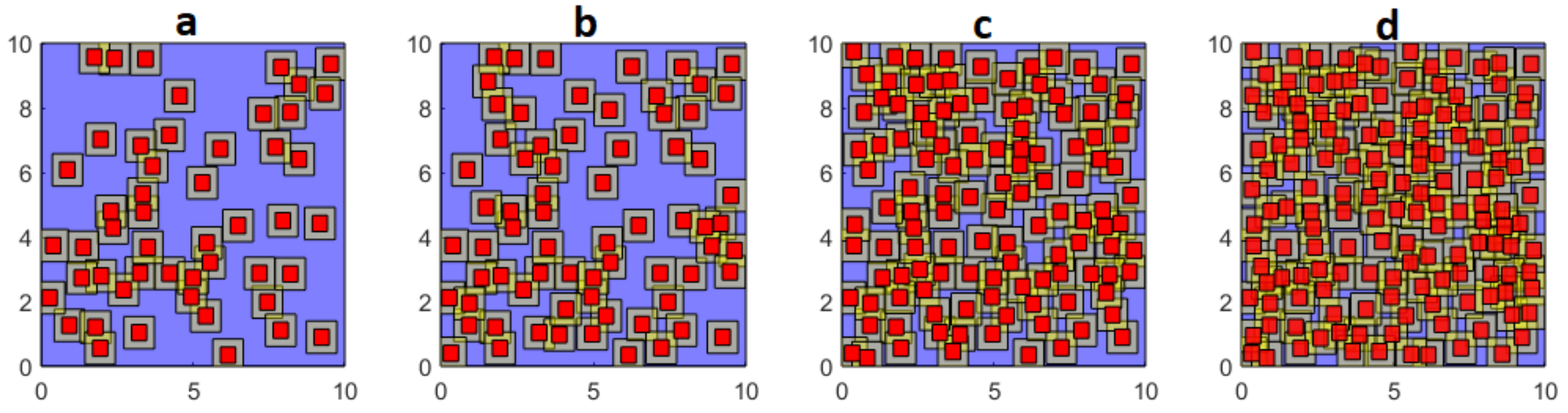

**Figure 3.** The process for creating two-dimensional square RSA packing is depicted in the image. The yellow parts denote locations where it would be hard to position the center of another square without it overlapping with the current ones, while the red squares show non-penetrable particles. The regions where the following square can be placed are shown by the purple voxels (shown without a grid). Each subsequent stage further subdivides these voxels into smaller, two-dimensional sub-voxels, which are then filled with fresh squares to represent the available space more accurately. Due to the lack of available space for purple voxels, the packing is virtually saturated in panel (d). Oriented square deposits are made.

### 2.2 Other viable algorithms

Aside from the SAT technique, straightforward tests utilizing their inscribed circles and circumscribed circles may also be used to determine if there is an intersection between two polygons: The two polygons must overlap if the inscribed circles do. The two polygons cannot overlap if the circumscribed circles do not. If these two simple tests fail to find a definitive answer, we then test if any two sides of the two polygons intersect (Zhang, 2018). An example of the application of this algorithm has been described in the reference (Zhang, 2018).

To calculate the inscribed radius and circumscribe radius of a polygon the following relationship has been taken advantage off:

$$R = a/2 * \sin\left(\frac{\mathrm{pi}}{\mathrm{n}}\right) \quad eq.1$$

$$r = R * \cos\left(\frac{\mathrm{pi}}{\mathrm{n}}\right) \quad eq.2$$

In which a is the side length of polygons, r inscribed radius and R is the circumscribed radius.

## 3. Results and discussion

To demonstrate the accuracy and usefulness of our algorithm, we build saturated RSA configurations of regular polygons and contrast saturation packing with prior findings documented in the literature; notably in the reference (Zhang, 2018); in which authors used a different algorithm and orientation were random. For each particle shape, we create 1000 permutations.

In (Zhang, 2018) the author describes a method for testing voxel availability in three-dimensional space for a particle simulation. The voxel space is represented by $(x, y, \omega)$, where $(x, y)$ is the location of the center of a polygon, and $\omega$ is the angle between the orientation of the polygon and some reference orientation. The voxel is defined as $(x \pm \delta x, y \pm \delta y, \omega \pm \delta \omega)$, where $\delta x$, $\delta y$, and $\delta \omega$ are half of the side length of the voxel in each direction.

To test if a voxel is available, a rigorous worst-case error analysis is performed to prove that no matter how x, y, and $\omega$ vary in the ranges $[x - \delta x, x + \delta x]$, $[y - \delta y, y + \delta y]$, and $[\omega - \delta \omega, \omega + \delta \omega]$, the upcoming particle will always overlap with an existing one. This is achieved by calculating the error bounds for the Cartesian coordinates of the vertices of the upcoming polygon inside the voxel. The method allows for the efficient testing of voxel availability in three-dimensional space and is suitable for use in particle simulations. The errors in the calculations could be small, which ensures accurate testing of voxel availability.

Nevertheless, while our technique (Zhang, 2018)'s for creating strictly saturated packings are comparable in terms of voxel removal, their respective bases for elimination are different.

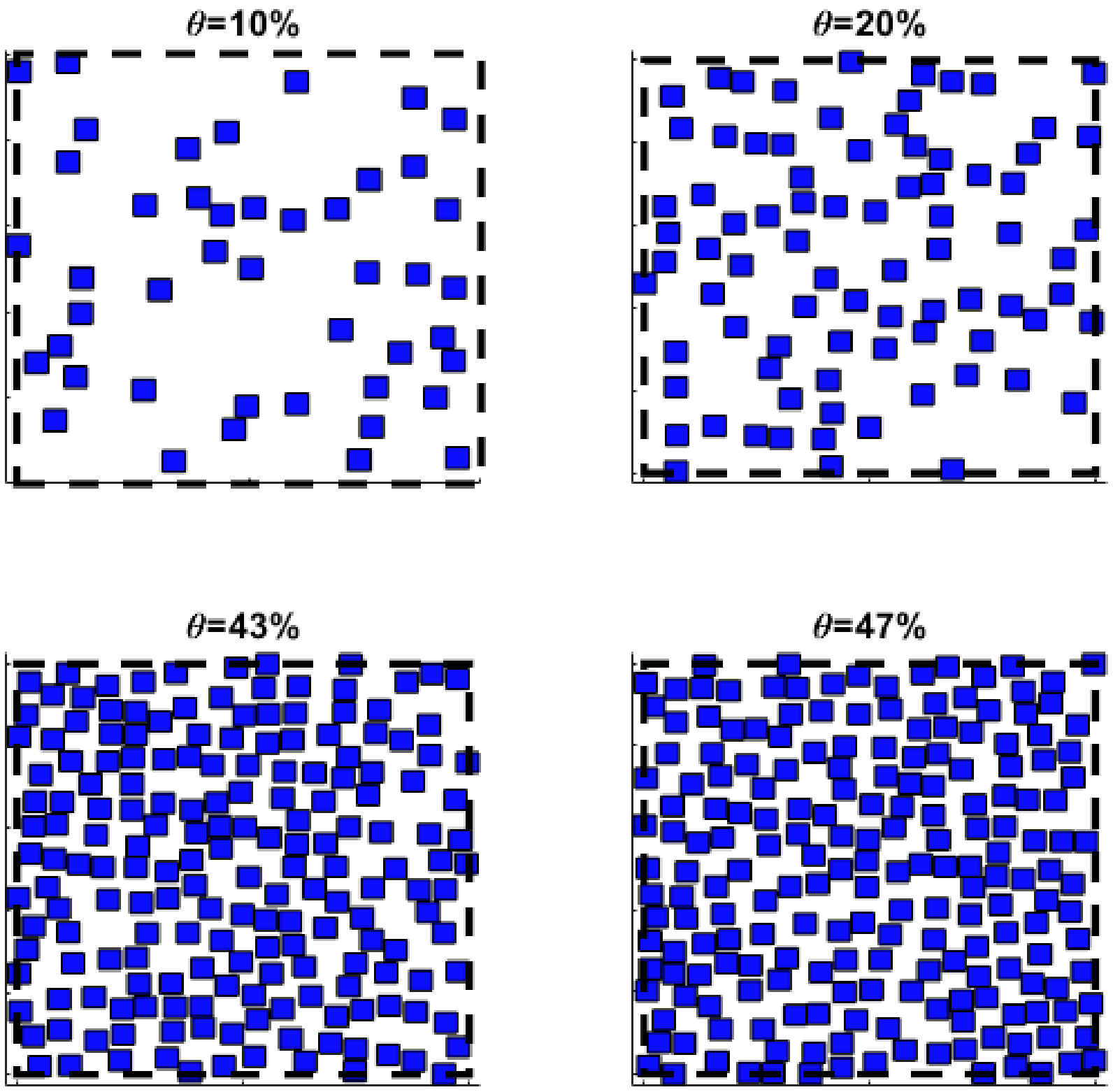

**Figure 4.** Typical monolayer samples for three different coverages: θ = 0.1, θ = 0.2, θ = 0.43, and θ = 0.47 for squares. The collector side length was equal to 15[-]. Fixed boundary conditions were used. Collector sides were chosen as 15 for better visibility of changes within RSA adsorption. Square deposits are oriented.

Most of the results discussed later in the paper were obtained using the largest area (50 by 50) with fixed boundaries; it is noteworthy that the side length of polygons was kept constant at 0.5 regardless. We checked that use of periodic boundary conditions does not have any significant influence on the presented conclusions. For instance, as will be revealed later, for the discs we arrived at a mean saturation packing value of 0.534803789±0.002720619 for fixed boundary condition at a ratio of disc surface area to collector surface area of 0.01; keeping the same dimension for periodic boundary condition leads to arriving at mean saturation packing of 0.546637121±0.000008419; thus, results here, while slightly different, can be compared later with studies conducting similar analysis using periodic boundary conditions.

To demonstrate visually how RSA works; the results of one of the simulations are depicted in **Figure 4** in which squares are deposited in an area of 15 by 15 with the square with a side of 0.5 and simulation is shown as saturation coverage gradually increases; in other words as simulation proceeds gradually more and more particles are deposited on the surface.

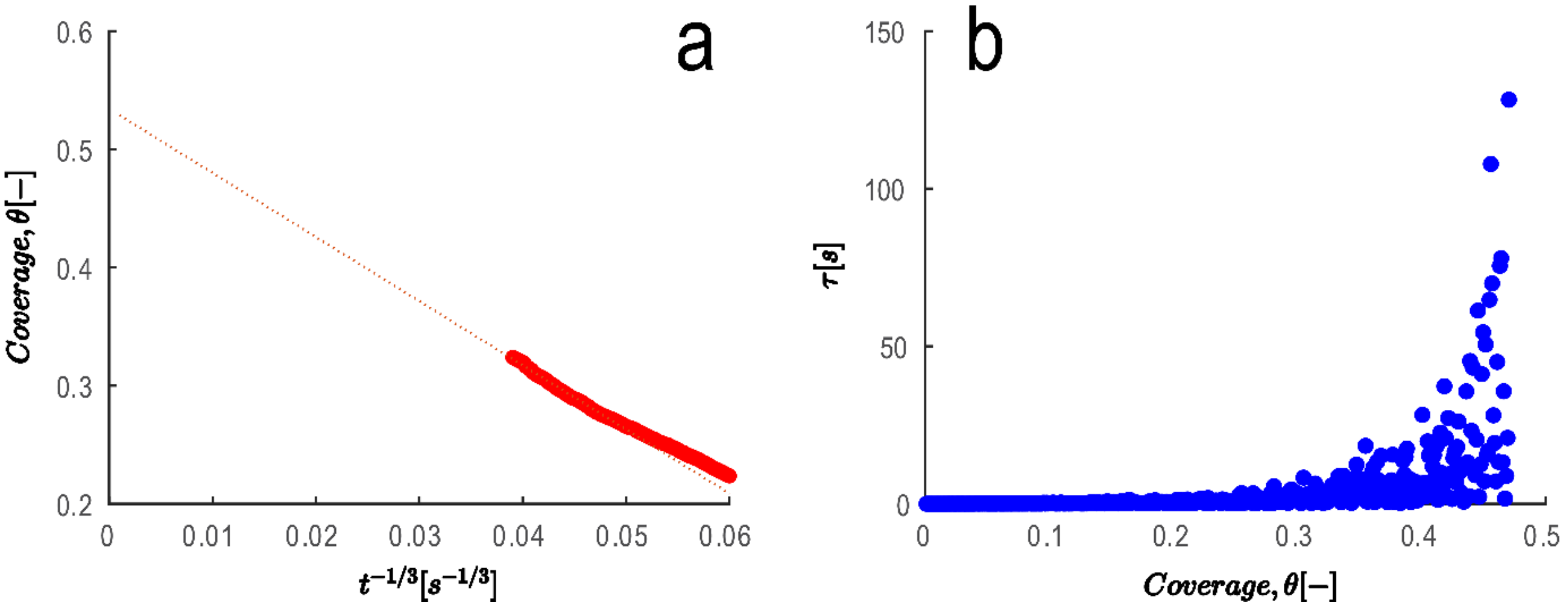

**Figure 5.** Squares with sides of 0.5 units put irrevocably inside a 15 by 15 space are the focus of the RSA algorithm. **(a)** Asymptotic observation of coverage as a function of total simulation duration indicated by t. **(b)** The instantaneous time τ determined based on coverage. Fixed boundary condition has been applied.

To look at the development in more detail, surface coverage was depicted as a function of simulation time to the power of $t^{-1/3}$ and results are shown in **Figure 5a**. Clearly at long simulation times of ~$10^4$, coverage very slowly reaches its asymptotic limit which is 0.5342139±0.00008. This asymptotic limit is estimated analytically, however, the technique presented below may also be readily followed to generate a comparable figure. This surface coverage corresponds very well with the results reported in the literature for squares (Akeda & Hori, 1976; Blaisdell & Solomon, 1970; Brosilow et al., 1991; Dickman et al., 1991; Feder & Giaever, 1980; Jodrey & Tory, 1980; Tory et al., 1983; Wang, 2000).

The number of attempts needed to add a new particle to the grid can be known, and this information can be used to model the blocking of further adsorption through monitoring time. As more of the surface is covered, adding new particles should be more difficult, which can be represented by a lower probability (See **Figure 5b**). Adsorption kinetics in a real experiment typically depends on two variables: the effectiveness of the transport process (primarily diffusion or convection depending on the experimental setup) that moves the adsorbate from the bulk to the surface, and the likelihood of catching particles that are nearby (Adamczyk et al., 1992; Adamczyk & Weroński, 1996; Evans, 1993; Schaaf & Talbot, 1989; Talbot et al., 2000; Viot et al., 1992). Authors in other reports(Cieśla, 2014) have concentrated on the second aspect in this case, which is defined by the blocking function, also known as the available surface function. The simulation makes it simple to obtain it as a ratio of successful attempts to all RSA attempts. Equivalent to the available surface function that is represented in the shape of time simulation is presented in **Figure 5b**. **Figure 5b** shows simulation time as a function of coverage; statistically, it is evident that more trials are required to attain adsorption because the surface is already rather packed.

Determining the mean saturation packing for squares and other polygons using the new algorithm and comparing it to the findings for hard discs were the major goals of this study. For an unlimited grid area and infinite adsorption period, that ratio should be given. Despite having to cope with finite simulation timeframes, one must control the inaccuracy caused by the finite grid size. The calculation of maximal coverage depends on the RSA kinetics model since it is uncertain if there will be any potential for adsorption beyond the simulation time, especially in the case of large grids. There were several works in the area, because of earlier studies in this field (Hinrichsen et al., 1986; Privman et al., 1991; Swendsen, 1981), it can be said that asymptotically:

$$\theta_{max} - \theta(t) \sim t^{-p} \quad eq.3$$

For irreversible deposition of disks or oriented squares; in which $\theta_{max}$ mean saturation packing and $\theta(t)$ is packing density at time, t.

One sample of the Results of fitting **Equation 3** was presented earlier in **Figure 5a**. It appears that the asymptotic observation of coverage for squares follows **Equation 3** nicely; therefore, the symptomatic simulation adopted here follows the previous finding correctly. Although no definitive proof of **Equation 3** has been given, analytical and numerical arguments have been presented that strongly support its validity(Viot & Tarjus, 1990). Note that for isotropic objects, p is equal to the number of dimensions, and **Equation 3** reduces to the usual Feder's law(Feder & Giaever, 1980). This study compares the outcomes of extrapolation experiments utilizing findings from previously published data in the literature to mean saturation packings obtained here using our newly created technique.

### 3.1 RSA for polygons

We built the groundwork for utilizing the RSA technique to produce oriented squares in the previous section as one of the polygons under investigation here. A sample of saturation densities for different shapes is presented in **Figure 6**. With the specific orientation used in this paper, as seen in **Figure 6**, hexagons are projected to have a high packing density due to their sides being well-suited for filling up the space in two dimensions. On the other hand, Pentagons with the fixed orientation adopted here are predicted to yield worse packings. Many methods have been developed throughout time to assess saturation packing, including extrapolation-based estimate (Viot & Tarjus, 1990) and more modern estimation based on recently developed algorithms that address some of the drawbacks of extrapolation methods(Cieśla et al., 2021; Zhang, 2018; Zhang & Torquato, 2013).

Extrapolation-based techniques use the following equation to estimate saturation packing:

$$\theta(t) = \theta_{max} + b/t^{p} \quad eq.4$$

In which, $\theta_{max} \equiv \theta\ (t \rightarrow \infty)$ and **Equation 4** is an extension of **Equation 3**. In estimating saturation packing, one gives the data from longer simulation times more weight when attempting to estimate $\theta_{max}$. If given all values the same weight, the approach's potential drawback is that each data point is given the same weighting factor. The higher section of the asymptotic area is sort of underweighted because there are a lot more of these points there. So, authors generally explore a new approach that creates a bias preferring the longest periods. These developments are consistent with the reports in the reference (Viot & Tarjus, 1990). The second method is to use algorithms that generate saturated packing these methods assist RSA to find empty areas more easily thus shortening the time of saturation considerably. The result of this approach will be given in the next section.

Direct applications of the simulation based on the shape discussed here are in chemistry and adsorption. For example, there are applications for water purification where a polymer globe must adhere to the particles, and "Pickering emulsion" applications where the peculiarly shaped kaolinite particles(Abbasi Moud & Hatzikiriakos, 2022; Abbasi Moud et al., 2021a) approach the interface and frequently form a monolayer adhesion  (de Folter et al., 2014) on the surface. In chemistry also, for instance, the form of compounds with five atoms, groups of atoms, or ligands organized around a central atom, defining the vertices of a pentagon, is referred to as pentagonal planar molecular geometry in chemistry(Housecroft & Sharpe, 2004).

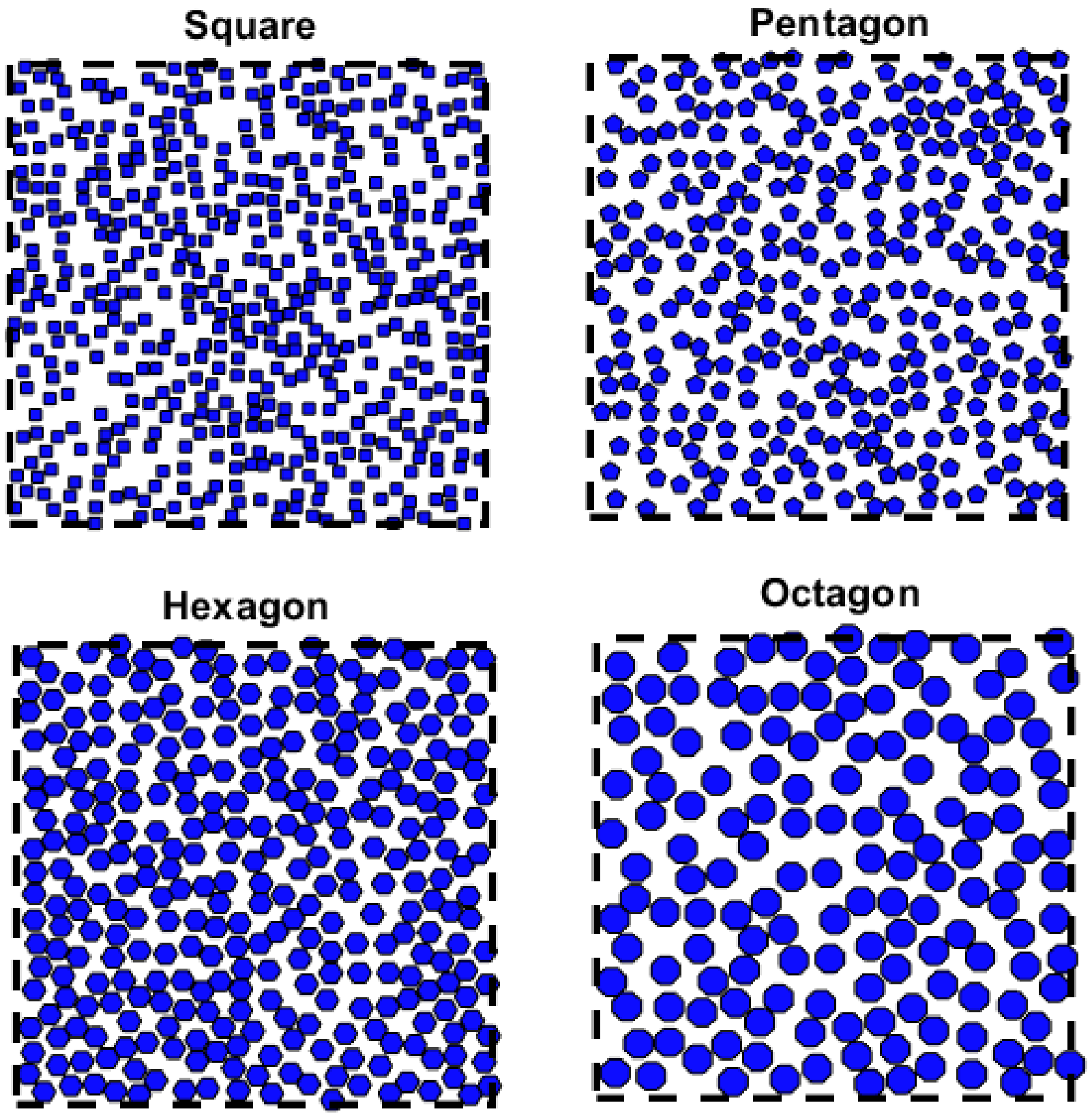

**Figure 6**. Square, pentagon, hexagon, and octagon near saturation points for samples with a side length of 0.5 and distributed within an area of 50 by 50. Fixed boundary condition has been applied.

**3.2 Voxel removal in numerical simulation**

Previously algorithm was left to run till the point that the addition of the next particle to the adsorbed mixed set of the particles in 2-d space became extremely sluggish; authors in the past, after arriving at this stage due to computational challenges, equation 4 would be used to estimate mean saturation packing; however, due to the nature of fitting and the dearth of a large number of points close to saturation, this approach would result in a slight overestimation of the true value of saturation packing. Here an algorithm inspired by the algorithm introduced earlier in the reference (Zhang, 2018) has been developed to precisely measure saturation packing through the voxel elimination method. For start, different shapes were placed on a square collector using the RSA protocol, with the surface area of the collector being 10000 times larger than the surface area of a single monomer. For each shape, we generated up to 1000 independent random packings to calculate the mean packing fraction, thus our results reached a statistical error of less than $2.0 \times 10^{-3}$. This allowed us to compare the effect of the polygon shape on the mean packing. To eliminate inactive voxels, we used the method described earlier; details of this methodology were explained in **section 2**. The parameter for triggering voxel division was determined based on the number of consecutive iterations without the addition of any particles. The optimal value of this parameter depends on the packing size, the shape of the deposited objects, and the maximum number of voxels, and is therefore difficult to estimate in advance.

To make the algorithm automatic a number is defined for consecutive iterations without adding particles that propel the algorithm to go through further voxel division. The optimal value of this parameter hinges on packing size, and the shape of the polygon deposited thus predicting this number prior is a challenging task and must be tweaked case by case. To give an example we found that for the objects studied here, the optimal value of this parameter can range from $10^2$ to $10^6$.

**Table 1.** Mean saturated packing density for regular polygons examined in the current manuscript using voxel-elimination methodology. The ratio of the polygon side to the collector side is 0.01.

| Objects | Mean saturation packing $\theta_{max}$ [-] | Standard deviation, δ [-] |
|---|---|---|
| **Squares** | 0.553933346 | 0.002713374 |
| **Pentagon** | 0.474351331 | 0.009773309 |
| **Hexagon** | 0.542599947 | 0.002987787 |
| **Heptagon** | 0.510433501 | 0.003200228 |
| **Octagon** | 0.527101701 | 0.001777729 |
| **Nonagon** | 0.513079313 | 0.003709104 |
| **Decagon** | 0.525761933 | 0.002968893 |
| **Dodecagon** | 0.529015894 | 0.001679424 |
| **Pentadecagon** | 0.516023238 | 0.004042373 |
| **Icosagon** | 0.514478961 | 0.009730145 |

**Table 1** presents the mean and standard deviation of the mean saturation packing for each type of polygon. The shapes of the polygons that are under consideration are listed in the "Objects" column. Squares, pentagons, hexagons, heptagons, octagons, nonagons, decagons, dodecagons, 15-gons, and 20-gons are among the forms included in the table. For squares, the mean $\theta_{max}$ is the highest, it is generally evident that as the polygon's number of sides increases, the mean decreases initially and then increases; however, it remains less than the mean saturation packing values for discs. In the same collector area, we also tried our voxel elimination approach for discs with a radius of 0.5, and we attained a mean saturation packing of 0.534803789±0.002720619.

Saturation coverage for equilateral triangles as 0.52590(Zhang, 2018), squares 0.523-0.532(Vigil & Ziff, 1989), 0.530(Viot & Tarjus, 1990), 0.530(Viot et al., 1992) and 0.52760(Zhang, 2018), regular pentagons 0.54130(Zhang, 2018), regular hexagons 0.53913(Zhang, 2018), regular heptagons 0.54210(Zhang, 2018), regular octagons 0.54238(Zhang, 2018), regular enneagons 0.54405(Zhang, 2018), and regular decagons 0.54421(Zhang, 2018) were previously for the same set of forms, although this time they were randomly oriented, were reported. The values reported here are like the values reported in these references however particles are oriented in the same direction as nematic structures in liquid crystals. As outlined in the introduction section, similarly for two dimensional aligned squares saturation coverage reported in the literature is 0.562009(Brosilow et al., 1991), 0.5623(Wang, 2000), 0.562(Feder & Giaever, 1980), 0.5565(Blaisdell & Solomon, 1970), 0.5625(Dickman et al., 1991), 0.5444(Tory et al., 1983), 0.5629 (Akeda & Hori, 1976), 0.562(Jodrey & Tory, 1980). Our square values are well within the range of values that have been published elsewhere. The boundary condition used here is fixed versus periodic variation in collector size ratio about polygon side, which may account for the modest discrepancies in values discovered between those provided. This information is reflected in **Table 2**.

**Table 2.** Demonstration of saturation coverage across literature; shapes are changeable from triangle to decagon with an arbitrary and fixed orientation.

| Objects | Mean saturation packing $\theta_{max}$ [-] |
|---|---|
| **Equilateral triangle** | 0.52590(Zhang, 2018) |
| **Square (unoriented)** | 0.523-0.532(Vigil & Ziff, 1989), 0.530(Viot & Tarjus, 1990), 0.530(Viot et al., 1992) 0.52760(Zhang, 2018) |
| **Square (oriented)** | 0.562009 (Brosilow et al., 1991), 0.5623(Wang, 2000), 0.562(Feder & Giaever, 1980), 0.5565(Blaisdell & Solomon, 1970), 0.5625(Dickman et al., 1991), 0.5444(Tory et al., 1983), 0.5629 (Akeda & Hori, 1976), 0.562(Jodrey & Tory, 1980) |
| **Regular pentagon** | 0.54130 (Zhang, 2018) |
| **Regular heptagon** | 0.54210 (Zhang, 2018) |
| **Regular octagon** | 0.54238 (Zhang, 2018) |
| **Regular enneagon** | 0.54405 (Zhang, 2018) |
| **Regular decagon** | 0.54421 (Zhang, 2018) |

As the shape of polygons progressively changes into discs as the number of sides rises, it is intuitive that the values provided in **Table 1** will drift towards the mean saturation packing of discs as the number of sides increases. Indeed, as we get closer to 12 sides, the difference between the dodecagon's claimed saturation coverage and that of the discs virtually disappears. We also observe that saturation is slightly lower for polygons with an odd number of sides than for their counterparts (pentagon and heptagon compared to hexagon and octagon); we hypothesize that this is because particles find it more challenging to pack efficiently because of the odd number of sides.

In conclusion, the densities of packings constructed of regular polygons are lower than the density of discs' packing for regular polygons explored here. Even side counts make packings denser because they reduce the average excluded area obstructed by a single shape. Increasing the number of sides aids to achieve more packed packings because it lowers the average excluded area blocked by a single shape. This area is measured by the parameter $B_2$ of the virial expansion(Baule & Makse, 2014; Ricci et al., 1992). Convex anisotropic figures, it has a simple form:

$$B_2 = 1 + \frac{P^2}{4\pi A_p} \quad eq.5$$

Where P is the perimeter and $A_p$ the surface area of the regular polygon. A lower value of $B_2$, where $B_2 = 2$ for a disc, can lead to denser packing at the beginning of the packing generation. However, elongated shapes may also form denser packings when aligned in parallel, especially close to the saturation limit where neighbors restrict possible orientations of subsequent objects. This competition (Cieśla et al., 2016; Haiduk et al., 2018; Vigil & Ziff, 1989) between the effects of $B_2$ and alignment can result in the densest RSA packings being observed for slightly elongated figures with small $B_2$ values and parallel alignments.

In three-dimensional packings, the highest packing fraction is often observed for moderately anisotropic ellipsoids(Donev et al., 2004; Man et al., 2005) or spherocylinders(Ferreiro-Córdova & Van Duijneveldt, 2014; Zhao et al., 2012). For set of objects considered here, the number is 2.273277 for squares that dwindles with the addition of sides to 2.008336 for Icosagon; thus, $B_2$ approaches the value of disk.

The results are not unexpected despite the reasons raised above since, if these polygons were stacked in a lattice form, one would find that squares should be the most tightly packed (due to a better chance of packing having no corners that could leave a portion of available space open). Additionally, since their neighboring polygons cannot occupy it, particles with odd numbers of sides will unquestionably have more space. The mean saturation packing for discs should be attained as the number of sides increases. Additionally, the form of a regular polygon might have an impact on its excluded volume. Because the hexagon is more rounded and has a bigger interior volume, it has a lower excluded volume than a normal pentagon of the same size.

### 3.3 Microstructure analysis

Details on the mean density of packed particles are contained in saturated random coverage. The density autocorrelation function may be researched to learn more about their composition. The radial distribution function (G(r)), also known as the pair correlation function, in a system of particles (such as atoms, molecules, colloids, etc.) explains how density changes in response to distance from a reference particle. G(r) is the radial distribution function (Hansen & McDonald, 2013) obtained from following equation:

$$G(r) = \frac{1}{\rho}\langle \sum_{i\neq 0} \delta(r - r_i) \rangle \; eq.6$$

The monolayer's first crucial characteristic is particle autocorrelation. **Figure 8** displays the average structures seen by various RSA packings to be precise we selected to compare the microstructure of packings at $\theta_{max}$ for square versus pentagons and square versus Icosagon. To make the graph comparable radius for the x-axis has been divided by the value of circumscribing radius described earlier in **Equation 1**. The peak in G(r) is growing for squares at the normalized radius of 1.414 for squares that are right on the edge of the particles; heralding that particles are neighboring each other closely. In other words, the function reaches its maximum for the closest neighbor, r = 0.5, and then begins to degrade because of the volume of particles that is lacking. A similar trend is seen in the pentagon and Icosagon, but there are more oscillating peaks. Indeed, the graph corresponding to Icosagon contains additional peaks, for instance, at r=2.5, 2.75, and 4.2, while the pentagons exhibit an additional stronger peak at ~3.9.

As the radius gets bigger, these peaks get weaker. Due to the coverage's randomness, these oscillations super exponentially vanish(Torquato et al., 2006), and after normalization, the function stabilizes at a value of 1. In addition, when the number of sides of polygons, the first peak corresponding to the nearest neighbor grows progressively sharper. According to this behavior, particles closer in shape to disks (Icosagon) pack more efficiently than squares.  Additionally, the sharpest peak for the set shown in **Figure 8** is sharper for Icosagons as opposed to Pentagons and squares.

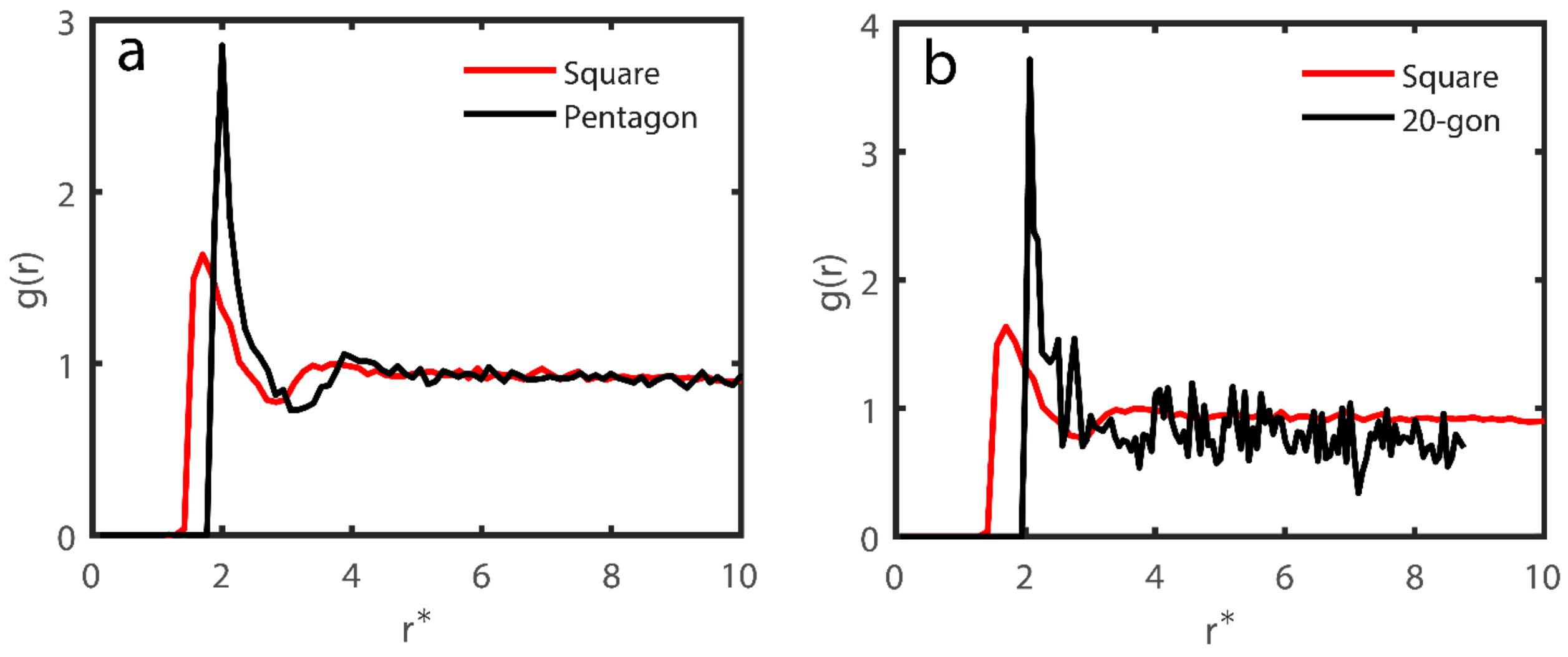

**Figure 7.** The normalized density autocorrelation function for selected regular polygons is depicted as a function of normalized radius for square, pentagons, and Icosagon.

To summarize, the average structure seen by a generic particle of the system described by G(r) displayed in **Figure 7**, shows a full agreement with the predicted theoretical regimes found in the literature (Donev et al., 2005; Richard et al., 1999). In all cases, we observe a pronounced peak at a distance r~0.5, with the sphere diameter that corresponds to the distance of the nearest neighbors in contact. For r larger than the diameter, the probability to find neighbors decreases. In fluid-like systems, theoretically for $\varphi \lesssim 0.55$, (Aste & Di Matteo, 2008; Donev et al., 2005; Richard et al., 1999) the G(r) is known to oscillate with decreasing amplitude something seen here.

### 3.4 Effects of finite sizes

It has been shown for RSA of disks that the measured value of saturated packing fraction for small packings oscillates concerning the system size, like the oscillations found in the tail of the autocorrelation function (Cieśla & Ziff, 2018). Therefore, it was expected (as a packing side length used in the study is much larger than 10 (normalized radius in **Figure 8**)) that the finite-size effects should not affect obtained values of packing fraction. To prove that the dependence of packing fraction on packing size for squares was analyzed and shown in **Figure 9**. Here we also present the packing fraction range reported earlier in references(Blaisdell & Solomon, 1970; Brosilow et al., 1991; Feder & Giaever, 1980; Wang, 2000), which is slightly higher but very close to the one reported here. This can happen with approximately one-third probability due to the probabilistic interpretation of standard deviations.

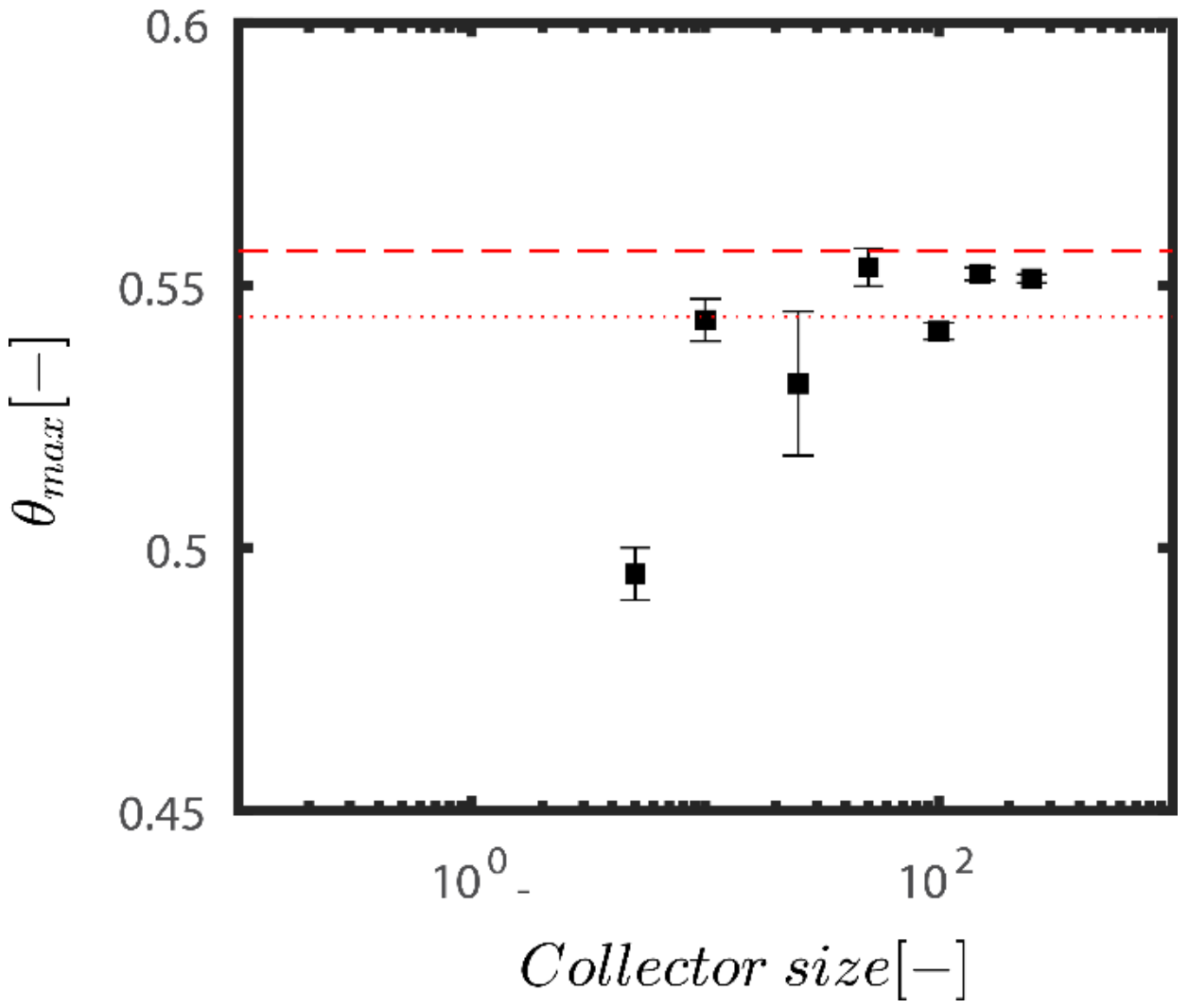

**Figure 8.** The relationship between the mean saturated packing and collector size is presented. For all packing sizes, the packing fraction was determined through the averaging of data collected from 100 independent random packings. The error bars indicate the standard deviation of the mean packing fraction. The two lines sketched corresponds to values reported for similar squares in references (Blaisdell & Solomon, 1970; Tory et al., 1983).

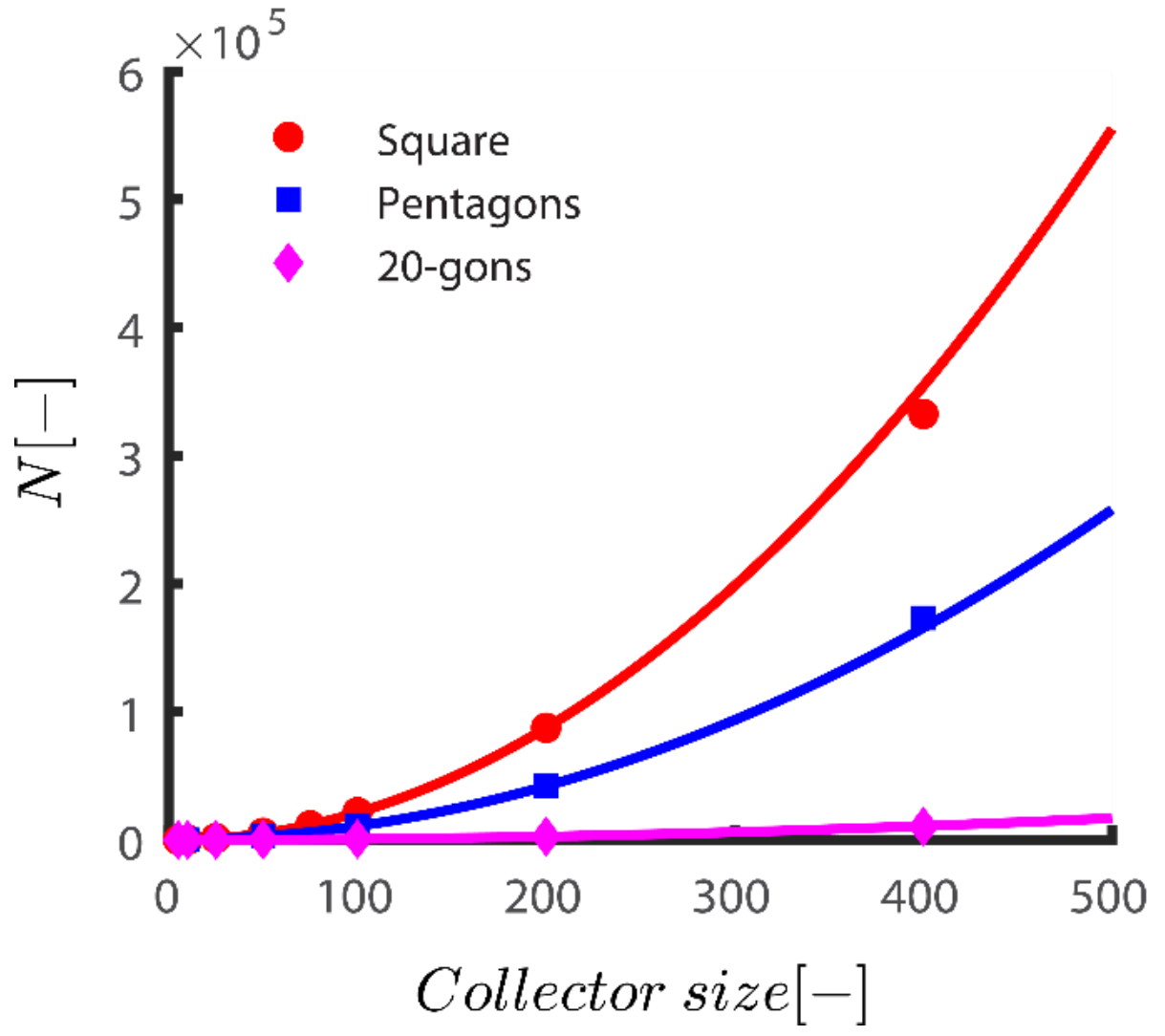

**Figure 9.** The dependence of the particle number in saturated random packing as the number of particles approaches infinity (N→∞) on the square surface side size (x = √collector size) for various exemplary shapes (Square, Pentagon, and Icosagon) is presented. The data represented by dots were obtained from numerical simulations, while the lines denote quadratic fits. The fits for each shape are as follows: N(square) = $142.4-11.96x + 2.244x^2$; N(pentagon) = $-73.33+7.36x + 1.017x^2$; N (Icosagon) = $1.321-0.2814x + 0.06867x^2$.

The same approach should be used on several systems of various, but finite sizes to ascertain the density of saturated packing on an arbitrarily large surface. **Figure 9** shows a quadratic relationship between the approximate number of packed particles and the square surface side length after an unlimited simulation duration. This behavior is comparable to that of randomly oriented regular polygons reported in reference (Cieśla & Barbasz, 2014).

## Conclusions

This study investigated the maximal random coverage or saturation coverage for a range of polygons, including squares through decagons, and compared the results against previous literature. The method used was a similar algorithm as those described in references (Cieśla et al., 2021; Zhang, 2018), which was modified and extended to find the RSA saturation densities of two-dimensional regular polygons with 4 to 20 sides, and verify their consistency with previous results reported in the literature. It is worth mentioning that these results apply to a wide range of particles including asphaltene, graphene, cellulose nanocrystals, and kaolinite among others(Abbasi Moud, 2022a; Abbasi Moud, 2022b; Abbasi Moud et al., 2021b). Although, it should be noted that due to the computationally challenging nature, it is not possible to say that all the configurations are global densities maxima, but it is believed that their densities are close to maximal.

The hope is that further research in this area would lead to the creation of new functional granular materials. The importance of this research is in the understanding of how tightly the particles that make up these materials are packed together can have an impact on their structure and physical characteristics. Examples of these materials include liquids, glasses, and granular substances, and the results suggest that by utilizing variously shaped particles, new materials with unique characteristics may be produced.

**Conflict of interest statement:** The author declares no conflict of interest.

**Data availability statement:** The datasets generated during and/or analyzed during the current study are not publicly available but are available from the corresponding author upon reasonable request.